# Elliptical instability of the flow in a rotating shell

Laurent Lacaze*, Patrice Le Gal, Stéphane Le Dizès

*Institut de Recherche sur les Phénomènes Hors Equilibre, 49 rue F. Joliot-Curie, BP 146,
Technopole de Château-Gombert, 13384 Marseille Cédex 13, France*

**Abstract**

A theoretical and experimental study of the spin-over mode induced by the elliptical instability of a flow contained in a slightly deformed rotating spherical shell is presented. This geometrical configuration mimics the liquid rotating cores of planets when deformed by tides coming from neighboring gravitational bodies. Theoretical estimations for the growth rates and for the non linear amplitude saturations of the unstable mode are obtained and compared to experimental data obtained from Laser Döppler anemometry measurements. Visualizations and descriptions of the various characteristics of the instability are given as functions of the flow parameters.

*Keywords:* Elliptical instability; Tidal deformations; Liquid core dynamics

## 1. Introduction

It is known from the seminal analysis of Kelvin (1880) that inertial waves whose origin comes from the restoring effect of the Coriolis force are eigenmodes of rotating fluid flows. These modes, neutral for inviscid flows but damped by viscosity, can however be observed in real flows when an external forcing is applied. For instance, Greenspan (1968) or McEwan (1970) visualized the wave patterns created by these inertial – or so called Kelvin – waves in rotating cylindrical tanks when a small excitation is added to the flow by the use of a small differentially rotating disc immersed in the fluid. Another example can be found in Aldridge and Toomre (1969) where some of the eigenmodes of a rotating sphere were excited by applying a resonant modulation on the rotating velocity. Later, using an inner deformable body placed inside a rotating sphere, Aldridge et al. (1997) and Seyed-Mahmoud et al. (2004) described some of the inertial waves of the shell with different frequencies and spatial structures. In this last experiment, some features of the elliptical instability were also detected. This instability arises from the resonant interaction of triads of waves: two Kelvin waves plus the elliptical deformation of the fluid streamlines by the boundaries (Waleffe, 1990). In fact, the elliptical instability has been intensively studied in the context of the transition to turbulence of shear flows (Widnall and Tsai, 1977; Widnall et al., 1974; Moore and Saffman, 1975; Pierrehumbert, 1986; Bayly, 1986; Bayly et al., 1988). There, the elliptical deformation of the streamlines inside cylindrical vortex cores are induced by other vortices (see for instance Leweke and Williamson, 1998a; Meunier and Leweke, 2001) or by mean shear fields (Le Dizès et al., 1996; Leweke and Williamson, 1998b; Lasheras and Choi, 1988). Contrary to these last studies relative to three-dimensional instabilities of rotating cylinders of fluids, the elliptical instability in spherical geometry rises completely different interests as it models the rotating inner liquid cores of planets subjected to tidal distortions induced by close gravitational bodies (Suess, 1971; Gledzer and Ponomarev, 1977; Kerswell, 1994). In particular the occurrence of this 'tidal' instability together with thermal or compositional convection in the molten cores of planets, such as the Earth, might be of prime importance in the generation or in the dynamics of the geomagnetic fields (Kerswell, 1994; Kerswell and Malkus, 1998). Recent measurements of magnetic fields around relatively small planets such as the Jovian moons Io and Ganymède (Kivelson et al., 1996a,b) may reinforce the interest in the study of inertial instabilities such as the elliptical or the precessional ones (Malkus, 1968; Busse, 1968; Noir et al., 2001; Kerswell, 1993; Lorenzani and Tilgner, 2003). Aldridge et al. (1997), and Seyed-Mahmoud et al. (2000, 2004) have performed computations and built as already mentioned, a rotating deformable shell where they observed the presence of the elliptical instability. Using the technique invented by Malkus in 1989 (Malkus, 1989), and more recently used and extended to triangular distortions by Eloy et al. (2003), we have applied an elliptical constraint to a deformable rotating sphere (Lacaze et al., 2004) and visualized the spin-over mode generated by the elliptical instability. This mode is a solid body rotation around an axis aligned along with the stretching direction. Moreover, from video image processing, we have measured the growth rates of the instability as functions of the flow parameters. These experimental results were finally advantageously compared to predictions resulting from theoretical linear and non linear analyzes. However, as it is the case for the Earth, it may often be that, due to huge pressure forces, planet inner cores crystallize and leave liquid shells between the planet mantles and the solid inner cores. Therefore, it is of some importance to study the effect of inner rotating solid bodies in the development of the elliptical instability in spherical geometry. The present article concerns an extension of our previous work on the spin-over mode of the elliptical instability in a full sphere (Lacaze et al., 2004) to the case of a rotating shell. In a first part, we present our theoretical model of the flow contained in the shell. The analysis of the elliptical instability in the inviscid limit and then in the viscous case, leads to the determination of the growth rate of the spin-over mode. In a second part, experimental results are presented. We used the same technique as before (Lacaze et al., 2004) but this time a solid small inner sphere is suspended by a thin wire in the center of a hollow transparent deformable external sphere. The ratio $\eta$ between both sphere radius has been chosen equal to 1/3 that is close to the value encountered in the Earth core. Visualizations and measurements by laser Döppler anemometry of several characteristics (growth rates, non linear saturation amplitude, spin-down time) of the spin-over mode of the shell are presented and compared to our theoretical predictions.

\* Corresponding author.
  *E-mail address:* lacaze@irphe.univ-mrs.fr (L. Lacaze); patrice.legal@irphe.univ-mrs.fr (P. Le Gal).

## 2. Linear stability analysis

### 2.1. Inviscid theory

The stability analysis of an inviscid fluid contained in a rotating and slightly deformed spherical shell is considered in this first section. It is assumed that only the outer sphere is deformed while the inner solid body remains spherical. Moreover, it is assumed that both spheres rotate around a vertical axis $e_z$ together at the same rate. This hypothesis is in accordance with our experimental device which will be presented later and is also a good approximation as regards to eventual geophysical applications. The elliptical deformation is considered to be a small perturbation amplitude parameter $\varepsilon$ ($\varepsilon \ll 1$).

The zero order problem, $\varepsilon = 0$, corresponds to the case of a rotating fluid in a non-deformed spherical shell for which the main flow can be written in cylindrical coordinates as:

$$\boldsymbol{U} = r\boldsymbol{e}_\theta, \tag{2.1}$$

which satisfies both boundary conditions

$$\begin{aligned} U = R_{i_H(z)} e_\theta &\quad \text{at } r = R_{i_H(z)}, \\ U = R_{o_H(z)} e_\theta &\quad \text{at } r = R_{o_H(z)}, \end{aligned} \qquad (2.2)$$

where $R_{i_H}(z) = \sqrt{R_i^2 - z^2}$ and $R_{o_H}(z) = \sqrt{R_o^2 - z^2}$ define the spherical boundaries in cylindrical coordinates. The variables have been non-dimensionalized with the angular velocity of the fluid $\Omega$ and the distance $d$ between the two shells. $R_i$ and $R_o$ are respectively the inner and outer sphere radii and their ratio is given by $\eta = R_i/R_o$.

The derivation of inviscid normal modes for the basic elliptical flow is not as trivial as it is in the case of the full sphere (Rieutord et al., 2001). For general rotating flows, the linearized Euler equations can be reduced to the so-called Poincaré equation (Greenspan, 1968). This equation is hyperbolic and its solutions are thus sensitive to the applied boundary conditions. By chance, in the case of the full sphere, a separation of variables can be achieved which permits to obtain the normal modes as explained in Greenspan (1968). The new variables are determined with respect to the outer surface and are thus no more consistent when an inner spherical body is added. To the best of our knowledge, no separation of variables have been discovered in the case of the shell. Rieutord and Valdettaro (1997) and Rieutord et al. (2001) studied this problem by considering the flow evolution along the characteristics of the hyperbolic equation. By this means, they have been able to obtain an approximate description of the inertial modes of the shell. In particular they show that most of them exhibit a complex spatial structure involving inner shear layers. However, as the spin-over mode is a solid body rotation whose axis is perpendicular to the main rotating flow axis, it represents a simple exact solution of the linearized Euler equations with free-slip boundary conditions on the inner and outer spheres (2.2) without inner shear layers. Viscous damping is therefore expected to be mostly due to viscous boundary layers as for the sphere (Hollerbach and Kerswell, 1995). This spin-over mode was found to be the most unstable mode in a deformed sphere for our experimental range of parameters (Lacaze et al., 2004). Thus it is reasonable to expect this mode to be also the most unstable in the shell. The experiment of Seyed-Mahmoud et al. (2004) and our own experimental observations presented in the next section confirm this expectation. Moreover, as in Lacaze et al. (2004), the inner shear layers induced by the boundary layer eruptions at the critical latitude (Hollerbach and Kerswell, 1995) are expected not to significantly modify our results and will not be considered in the following.

The stability analysis of the elliptical instability of a deformed solid body rotation in a shell is considered when an order $\varepsilon$ deformation of the external boundary is imposed to the main flow defined by Eq. (2.1). The deformation, stationary and two-dimensional, has an azimuthal wavenumber equal to two. As already mentioned, the inner sphere remains spherical. Splitting the flow in three regions, an inviscid solution can thus be written in cylindrical coordinates as follows:

$$U = U_r e_r + U_\theta e_\theta, \qquad (2.3)$$

$$\left|\begin{aligned} U_r &= \varepsilon r \frac{R_{o_H}(z)^4}{R_{o_H}(z)^4 - R_{i_H}(z)^4} \times (1 - R_{i_H}(z)^4 r^{-4}) \cos(2\theta) \\ U_\theta &= r - \varepsilon r \frac{R_{o_H}(z)^4}{R_{o_H}(z)^4 - R_{i_H}(z)^4} \times (1 + R_{i_H}(z)^4 r^{-4}) \sin(2\theta) \end{aligned}\right. \quad \text{for} \quad -R_i < z < R_i,$$

$$\left|\begin{aligned} U_r &= \varepsilon r \cos(2\theta) \\ U_\theta &= r - \varepsilon r \sin(2\theta) \end{aligned}\right. \quad \text{for} \quad R_i < |z| < R_o.$$

In the two polar regions defined by $R_i < |z| < R_o$, this flow is equivalent to the main flow in a deformed rotating full sphere (Lacaze et al., 2004). Elsewhere, $-R_i < z < R_i$, and contrary to the model used in Seyed-Mahmoud et al. (2004), the inner sphere influences the main flow by inducing a potential flow of same order as the imposed deformation. This main flow defined by Eq. (2.3) is singular at the poles of the inner sphere but is regular and continuous elsewhere. Moreover, it satisfies the inviscid boundary condition of non penetration at both solid surfaces.

Iso-values of the azimuthal velocity $U_\theta$ in a meridional plane for $\eta = 1/3$ and $\eta = 3/5$ are presented in

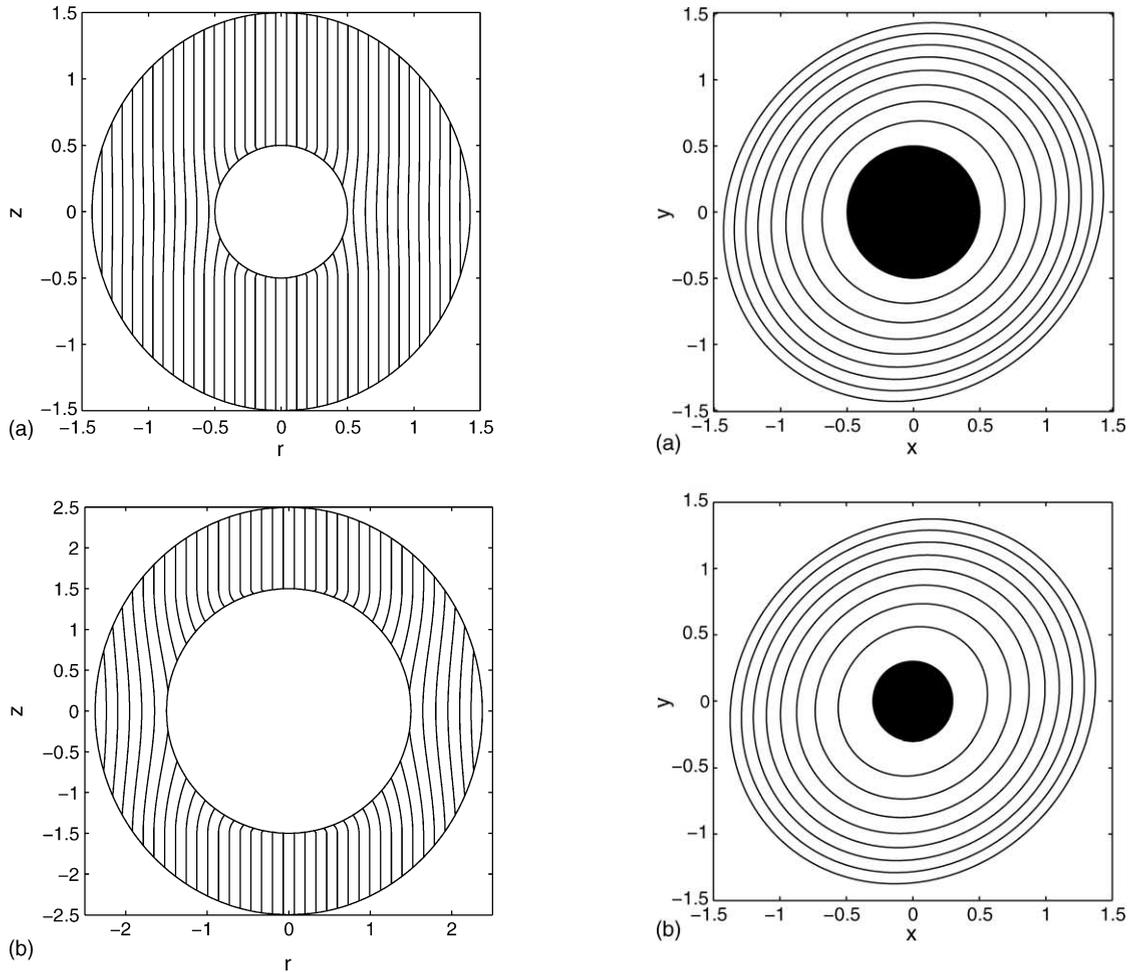

Fig. 1. Iso values of azimuthal velocity for $\eta = 1/3$ (a) and $\eta = 3/5$ (b). In both cases $\varepsilon = 0.125$.

Fig. 1. When the shell is not deformed or in the case of a deformed full sphere, these iso-values are vertical lines in meridional planes. As can be seen, the addition of the inner sphere slightly modifies the flow principally in the vicinity of the core. This modification is due to the potential flow added in Eq. (2.3). We observe that the larger is the inner sphere the more the main flow structure is deformed.

The streamlines of the main flow in horizontal planes corresponding to $z = 0$ (equator), $z = 0.4$ and $z = 1$ are shown in Fig. 2. They illustrate the basic flow patterns in the equatorial and polar regions as defined before. The black disc represents a horizontal section of the inner sphere. In Fig. 2a and b, it can be seen that the streamlines exhibit a different structure as that of Fig. 2c. For the two first cases, the elliptical streamline which defines the external boundary progressively becomes a circle close to the inner sphere contrary to the third figure where similarly to the case of the full sphere, the streamlines are concentric ellipses with the same eccentricity.

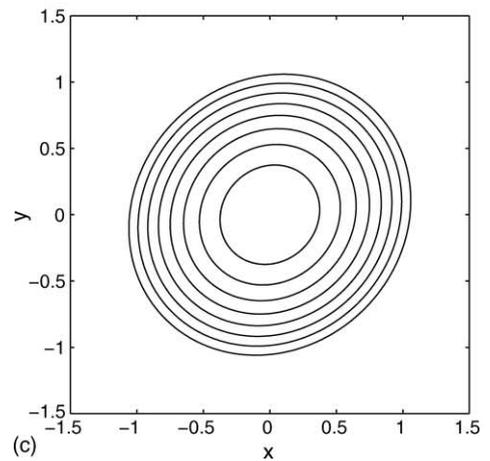

Fig. 2. Streamline function at $z = 0$ (a), $z = 0.4$ (b), $z = 1$ (c).

As in the case of the full sphere (Kerswell, 1994) or the cylinder (Malkus, 1989; Eloy et al., 2003), the order $\varepsilon$ flow induced by the elliptical deformation can resonate with two Kelvin modes of the spherical shell if they satisfy the resonant condition: $\omega_2 = \omega_1$ and $m_2 = m_1 \pm 2$ (($\omega_1, m_1$) and ($\omega_2, m_2$) are the frequencies and the azimuthal wave numbers associated with the two considered modes). The third condition specified by Kerswell (1994) in the case of the full sphere, which corresponds to a spatial coherence between each modes in a meridional plane (or the equivalent in a cylinder: $k_1 = k_2$ where $k_1$ and $k_2$ are the axial wave numbers of the two modes), is not well defined here due to the unknown form of the mode basis. However, as indicated above, we are interested in the peculiar case of the resonance of the spin-over mode which corresponds to a combination of the two inertial waves characterized by $(m_1, m_2) = (-1, 1)$ and $(\omega_1, \omega_2) = (0, 0)$ with the basic elliptical flow. This resonance leads to an order $\varepsilon$ exponential growth in time of the two symmetric modes, which can be written in cylindrical coordinates

$$\boldsymbol{u} = \mathrm{e}^{\pm i\theta} \begin{vmatrix} \mp iz \\ z \\ \pm ir \end{vmatrix} + \mathrm{C.C}$$

The derivation of the inviscid growth rate is classical and has already been done many times in different configurations (Moore and Saffman, 1975; Tsai and Widnall, 1976; Eloy et al., 2003; Lacaze et al., 2004). The method consists in applying a condition of solvability for the corrected flow at order $\varepsilon$. This condition of solvability is then solved by determining the adjoint modes which is in this case the mode itself corrected by a boundary condition term $\mathcal{I}$. The inviscid growth rate $\sigma_{NV}$ is then determined by resolving the relation of dispersion at order $\varepsilon$. This leads to the general growth rate expression:

$$\sigma_{NV} = \frac{\mathcal{N} - \mathcal{I}}{\mathcal{J}}, \tag{2.4}$$

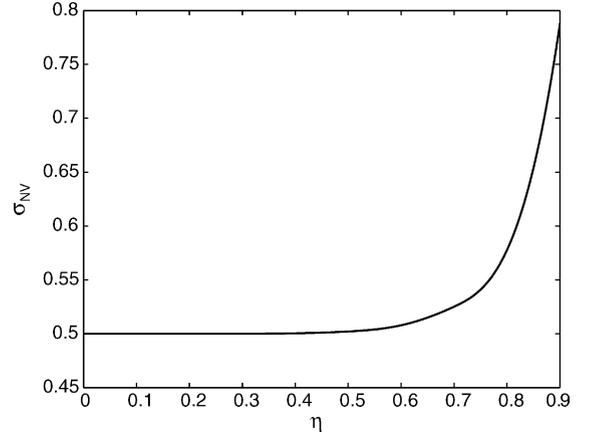

Fig. 3. Evolution of the spin-over growth rate with $\eta$ in the inviscid theory. For moderate $\eta$, no variation from the full sphere case is observed.

where $\mathcal{J}$ is the energy of the spin-over mode, $\mathcal{I}$ is the surface boundary condition term and $\mathcal{N}$ is the interaction of the spin-over mode with itself via the order $\varepsilon$ correction of the main flow (2.3). The parameter $\eta$ only enters Eq. (2.4) through the term $\mathcal{N}$ for $-R_i < z < R_i$. All the other terms are equivalent to that of the full sphere case.

We have calculated this inviscid growth rate $\sigma_{NV}$ for the spin-over mode. It is plotted in Fig. 3 as a function of $\eta$. We note that for $\eta = 1/3$, which is our experimental configuration, the growth rate is not significantly different from the inviscid growth rate $\sigma_{NV} = 1/2$ of the full sphere case. The same remark was also formulated in Seyed-Mahmoud et al. (2000) concerning the frequencies of the Kelvin waves in the sphere and in the $\eta = 1/3$ shell.

### 2.2. Viscous effect

In the previous section, the inviscid growth rate of the unstable spin-over mode has been determined. To allow further comparisons with the experimental results, we need to take into account the viscous dissipation terms (or at least the most significant ones). The ratio between the viscous forces and inertia is measured by the Ekman number $E = \nu/\Omega d^2$. The instability threshold should then be obtained as function of $E$ and $\eta$. As mentioned above, the flow described by Eq. (2.3) satisfies inviscid boundary conditions but not the no slip condition (or viscous boundary condition) in the region defined by $-R_i < z < R_i$. This implies the existence of Ekman viscous boundary layers both on the inner and outer shells for $-R_i < z < R_i$. It has already been shown that the Ekman layers have a thickness of order $E^{1/2}$ (Greenspan, 1968) with an order $\varepsilon$ correction flow that permits to satisfy the no slip condition. Therefore, when these Ekman layers are regular, the viscous correction to the flow implies an order $E^{1/2}$ damping and an order $\varepsilon E^{1/2}$ recirculation flow in the bulk, which can be neglected. Depending on the flow geometry, volume dissipation effects ($O(E)$) could also become important if the spatial structure of the mode is complex. For the cylinder, Eloy et al. (2003) demonstrated that these effects were responsible for the damping of large wavenumber modes. By contrast, for the sphere, Zhang et al. (2004) showed that volume dissipation effects were null for each mode, whatever complex its spatial structure. We do not know whether the shell satisfies the same property. But, in the present study, it is not our concern. We consider a single mode (the spin-over mode) for which the volume dissipation is identically zero.

Let us remark that spherical Ekman layers possess a singular behavior at the critical latitudes so that inner shear layers are spawned from these singular points and penetrate into the inviscid flow. We can also show that the viscous correction to the inviscid main flow (2.3) is also singular at the poles of the inner sphere. But as done in many linear studies, we will not consider here the dissipation associated with the regularization of these singularities as they are generally found to be weak (Hollerbach and Kerswell, 1995). However we should mention that these effects could become important in highly non linear regimes at low Ekman number.

These different assumptions permit to give a prediction on the threshold of the elliptical instability. As shown by Hollerbach and Kerswell (1995), the damping of the spin-over mode in a spherical shell is,

$$\sigma_V = 2.62 f(\eta)$$
$$\text{where } f(\eta) = \frac{(1+\eta^4)(1-\eta)}{1-\eta^5}. \quad (2.5)$$

The value 2.62 is the spin-over damping rate in the case of the full sphere and the function $f(\eta)$ is the correction induced by the increase of dissipation in the inner sphere boundary layer. Thus, the estimation of the viscous growth rate close to the threshold reads,

$$\sigma = \varepsilon \sigma_{NV} - E^{1/2} \sigma_V.$$

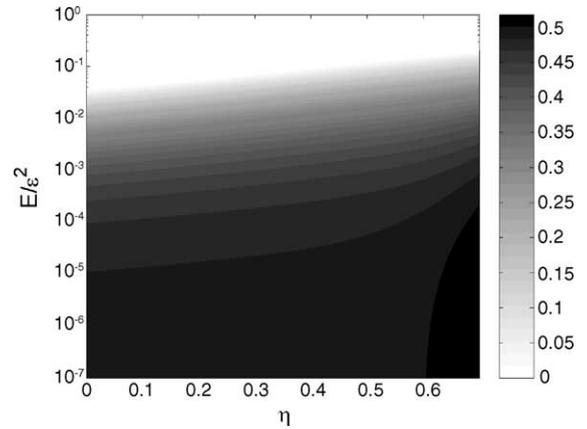

Fig. 4. Viscous growth rate as function of the Ekman number and the geometrical parameter $\eta$.

Fig. 4 shows the evolution of this spin-over mode viscous growth rate $\sigma/\varepsilon$ in the $(\eta, E)$ plane. For a given geometry, $\eta$ and $\varepsilon$ are prescribed and the evolution of $\sigma$ as function of $E$ is simply given by a vertical cut of Fig. 4. This kind of curves will be used for the forthcoming comparisons between theory and experiments.

## 3. Experimental results

### 3.1. Experimental techniques

The experimental device was already used by Lacaze et al. (2004) (see Fig. 5). The only modifications in the experimental arrangement come from our desire to build a hollow shell this time. For this purpose, a ping pong ball has been opened and a small solid ball introduced inside. This small sphere is rigidly mounted on a thin 0.2 mm in diameter nylon thread going through both spheres along their diameters. This thin wire is also used to position the inner ball in the middle of the hollow sphere. The ping pong ball is then closed back and polished to recover its perfect spherical shape. It is then inserted in a cylindrical block of liquid silicone that is cured at a temperature of 50 °C with the ping-pong ball inside. Finally, the ping-pong ball is dissolved by a solution of ethyl acetate and a hollow sphere molded in a transparent and deformable cylinder with a small core inside is obtained. The radii of the outer and inner sphere are respectively 21.75 mm and 7.5 mm. These values give a value of $\eta$ approximately equal to 1/3 which is in accordance with the geophysical situation relative to the Earth. The silicone cylinder is mounted on the vertical shaft of the device used in Lacaze et al. (2004) and is compressed between two vertical rollers. Note that these rollers are always in place; the device does not offer the possibility to move them in or out after rotation is started. The distance separating these rollers gives directly the elliptical deformation of the outer deformable sphere. Here, a value of $\varepsilon = 0.125$ is chosen. The study has been done in a range of angular velocity going from 0 rpm to 150 rpm so that the Ekman number varies from $3.14 \times 10^{-4}$ to $\infty$.

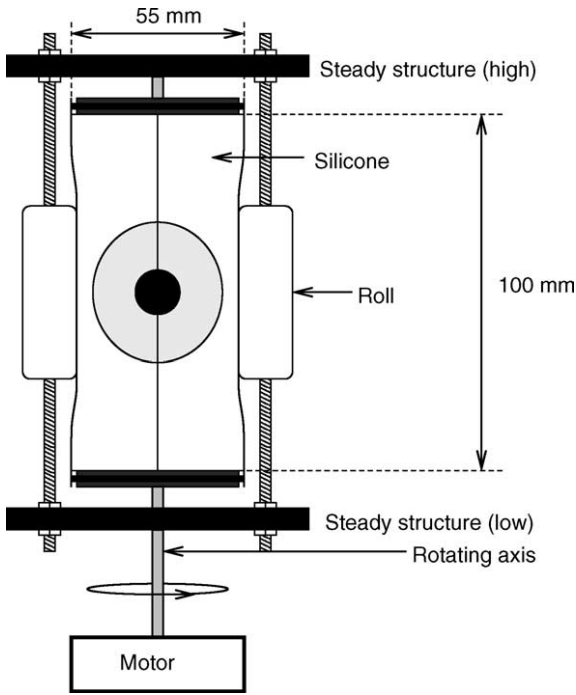

Fig. 5. Experimental device with the hollow sphere molded in an elastic and transparent cylindrical block of silicone gel and containing the inner small sphere. The silicone cylinder is compressed by two rollers as it rotates around its axis.

The flow is visualized using a meridian laser plane illuminating the sphere that is filled with water seeded by kalliroscope flakes. The elongated shape of these reflective particles allows them to align in the flow and visualize the velocity field. In particular, the rotation axis of the flow is clearly visible as shown on Figs. 7 and 8. As the spin-over mode is mainly a solid body rotation around an axis perpendicular to the entrainment rotation axis, the combination of the main rotation with this mode leads to an axis inclined at an intermediate angle. In Lacaze et al. (2004), the measurements of this angle in time permit the determination of the instability growth rate. Unfortunately, here, because of the presence of the inner sphere, this simple technique was not accurate enough and flow velocities have been measured by laser Döppler anemometry. This system can measure two projections of the velocity field in the vertical and azimuthal directions. For this purpose, the fluid has been seeded with spherical particles of diameter 10 μm. The two measurement directions of the velocity field are shown on Fig. 6. The measurement location point in the middle of the fluid shell is chosen at the intersection of the equatorial plane and the meridional plane perpendicular to the maximum stretching direction induced by the rollers (e.g. 45° from this one). On the line defined by the intersection of these two planes, the flow can be written as

$$\boldsymbol{U} = V_\theta \boldsymbol{e}_\theta + V_z \boldsymbol{e}_z,$$

where $V_\theta = r$ is the flow velocity induced by the rotation and $V_z = V_0 r e^{\varepsilon \sigma t}$ is the velocity associated with the exponentially growing spin-over mode. The initial amplitude $V_0$ that gives rise to the instability is in fact not controlled in the experiment. It simply comes from experimental fluctuations of the flow.

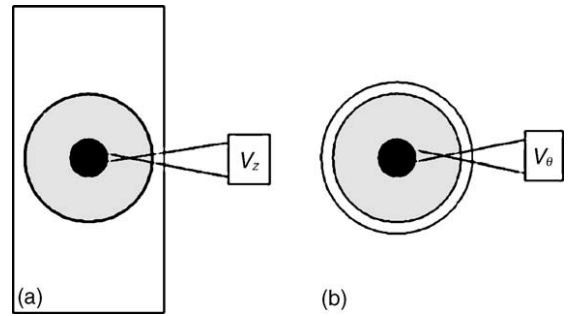

Fig. 6. Laser Döppler anemometry arrangement: (a) side view, (b) top view.

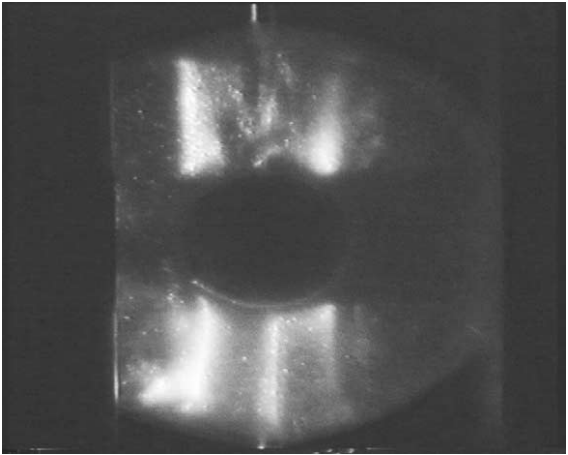

Fig. 7. Visualization of the spin-up phase after the start of the shell ($\eta = 1/3; \varepsilon = 0.125$).

## 3.2. Visualizations and measurements

Two snapshots of a typical evolution of the flow in our deformed spherical shell are presented in Figs. 7 and 8. The first figure presents an image of the flow during the spin-up transient. Inner cylindrical shear layers separating rotating fluid to steady fluid are particularly well visualized by the two vertical bright lines (intersection of a cylinder with the laser vertical plane) that propagate from the outer boundary towards the axis of rotation. After some minutes, this axis tilts as the flow gets unstable. As already mentioned, the flow is a combination of the main rotating flow around a vertical axis and the spin-over mode rotation whose axis is given by the stretching direction (at least for the considered range of Ekman numbers).

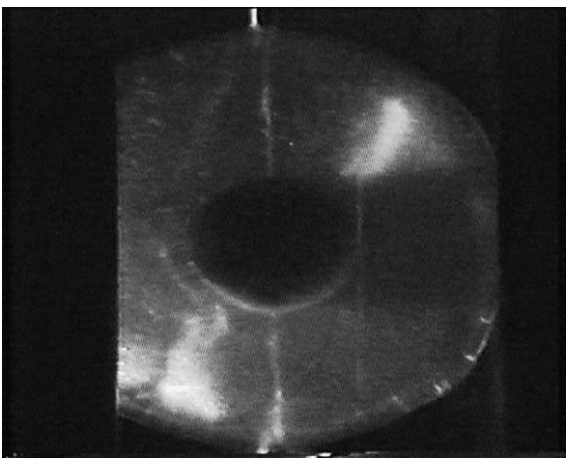

Fig. 8. Visualization of the spin-over mode of the elliptical instability in a slightly deformed rotating shell. $E = 4.7 \, 10^{-4}; \eta = 1/3; \varepsilon = 0.125$.

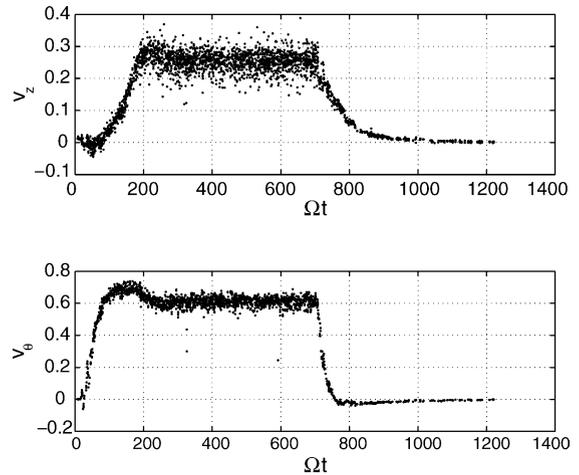

Fig. 9. Velocity time series for $E = 4.7 \, 10^{-4}; \eta = 1/3; \varepsilon = 0.125$. Top: axial velocity ($V_z$); bottom: azimuthal velocity ($V_\theta$).

These two snapshots show the strong similarity between the full sphere case studied in Lacaze et al. (2004) and the shell case. As already observed by Seyed-Mahmoud et al. (2004), the presence of the inner sphere does not strongly modify the structure of the unstable mode. To get a quantitative description of the instability, Laser Döppler anemometry time series are recorded and presented for $E = 4.7 \times 10^{-4}$ in Fig. 9. Velocities are non-dimensionalized with the maximum velocity of the outer spherical boundary along the equator. At $t = 0$, the device is set into rotation. As can be seen in Fig. 9, the azimuthal velocity first grows during the so called spin-up phase. Then, when the azimuthal velocity has saturated, the axial velocity which would remain null without any instability, grows exponentially till a saturated regime is reached. Later ($\Omega t = 700$), the motor is abruptly stopped and the spin-down phase is also recorded.

Although the spin up and spin-down regimes are well-known, experimental verifications of the scaling law are rare. In our experiments, scaling laws can be obtained from the azimuthal velocity data, from which we can estimate the duration of the spin-up phase by determining the time needed to reach the maximum velocity. This time which depends on the Ekman number, is plotted as a function of $E^{-1/2}$ in Fig. 10. This scaling has been proposed by Greenspan (1968) and the straight line which fits the data points, confirms this $E^{-1/2}$ behavior. For the lowest values of the Ekman number ($E^{-1/2} > 45$), a bending of the curve is observed which can be explained by the apparition of the instability before the end of the spin-up. In fact the non linear interaction between the main flow and the unstable mode generates a negative vertical vorticity (Kerswell, 2002; Lacaze et al., 2004) that invalidates our time estimation. In the same way, we will see in the following that the instability growth rate measurements are also affected by this interplay for the two smallest values of the Ekman number we have considered. Nevertheless, it is worth mentioning that for $E > 4.4 \times 10^{-4}$ ($E^{-1/2} < 45$), time scales associated with spin up and instability are well separated. For these Ekman numbers good estimates for the spin-up scaling laws and for the instability growth rate are obtained from the anemometric time series. The growth rate is extracted from the time series by plotting the data in a semi-logarithmic graph as illustrated in Fig. 11. In this figure, the linear fit which corresponds to the exponential growth of the instability before its non linear saturation, is superimposed on the experimental measurements. The slope of this straight line gives a direct measure of the linear growth rate of the instability. A systematic measure of this exponential growth rate versus the inverse Ekman number is given in Fig. 12. In this figure, the experimental measurements are compared with the results calculated from the linear analysis described in the previous section. The theoretical curve corresponds to an ellipticity $\varepsilon = 0.125$ and a ratio $\eta = 1/3$, and an excellent agreement between theory and experiment can be observed for most of the points. The error bars are estimated from the scatter in the slope measurements on the velocity time series. As expected and only for the largest Ekman numbers, the measure of the experimental growth rate deviates from the linear theory.

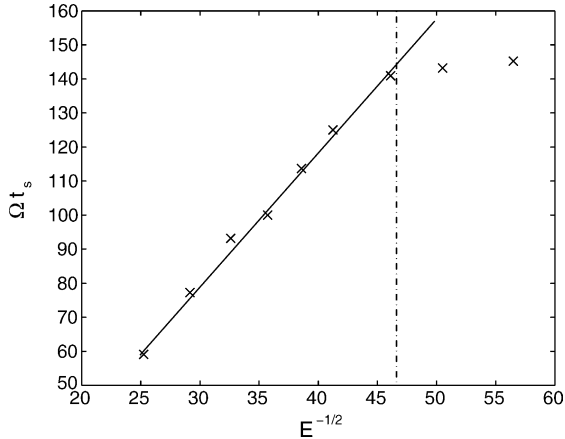

Fig. 10. Spin up characteristic time as function of $E^{-1/2}$ ($\eta = 1/3$; $\varepsilon = 0.125$).

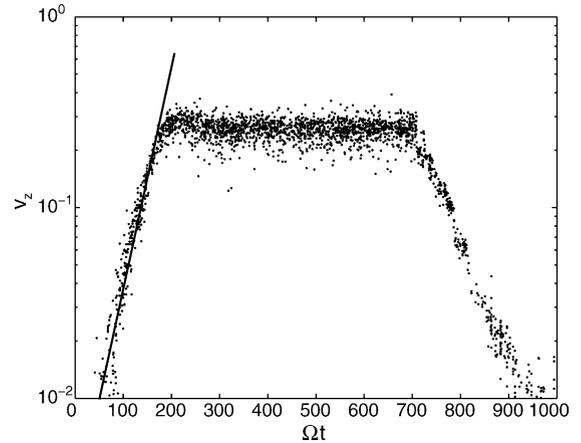

Fig. 11. Time series of the vertical velocity for $E = 4.7 \times 10^{-4}$; $\eta = 1/3$; $\varepsilon = 0.125$.

After the linear growth phase, the spin-over mode saturates and the flow is a steady tilted rotation which is visualized on Fig. 8. A slight overshoot which comes from the non linear interaction between the mode and the main rotation (at high amplitude, the unstable mode decreases the amplitude of the main rotation), is visible on the velocity time series. The asymptotic saturation amplitude of the instability is measured through the saturation of the vertical velocity at large time. Both saturation and maximum amplitudes are represented in Fig. 13. In this figure, the measures are also compared with the theoretical predictions obtained from the non linear model used for the sphere (Lacaze et al., 2004). This non linear model has been adapted to the shell case by multiplying all viscous terms by the function $f(\eta)$ given in (2.5). The simulation of this model leads, as it was the case for the full sphere, to a saturation of the spin-over mode. As already noted, the shell case for $\eta = 1/3$ and the hollow sphere case are similar. It can be verified that the experiment and the theoretical analysis for the shell lead to comparable results which are presented in Fig. 13. The major effect of the presence of the inner core is to decrease the saturation amplitude of the mode. As in the experiments, the model exhibits an overshoot (solid line) that can be measured and compared to the experimental points (×). In the same way, a good agreement is found between the asymptotic (large time) values obtained from the model (dotted line) and from the experiment (+).

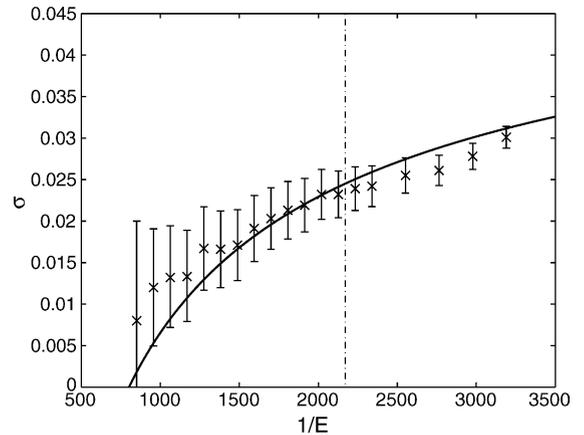

Fig. 12. Comparison between measurements of the experimental growth rates and their theoretical predictions ($\eta = 1/3$; $\varepsilon = 0.125$).

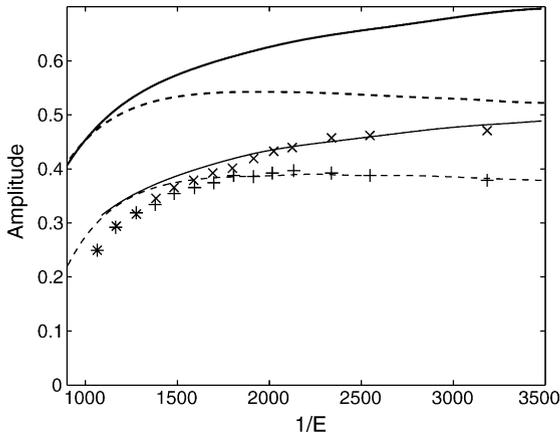

Fig. 13. Asymptotic (dotted line) and overshoot maximum of amplitude (solid line) from numerical simulation of the instability model. The heavy curves are computed for the full sphere case and the others for the $\eta = 1/3$ shell ($\varepsilon = 0.125$).

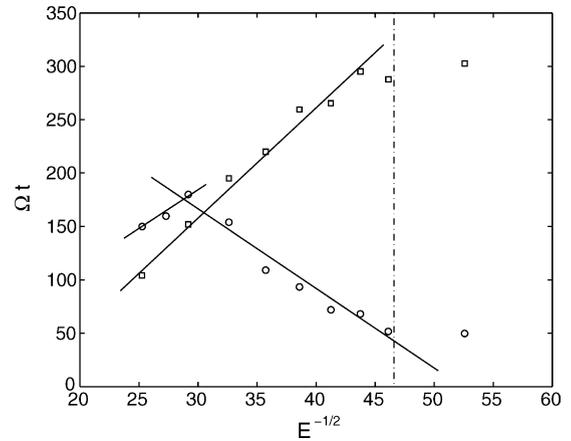

Fig. 14. Damping characteristic time of the spin-over mode, measured from the vertical velocity (squares), damping time of the main rotation measured from the azimuthal velocity (circles) as functions of $E^{-1/2}$ ($\eta = 1/3; \varepsilon = 0.125$).

When the rotation of the outer sphere is suddenly stopped, the flow velocity decreases because of viscous damping. As can be seen on the right part of the curve presented in Fig. 11, an exponential decay is observed. Both the characteristic damping times of the azimuthal velocity (associated with the main rotation) and of the axial velocity (associated with the spin-over mode) are plotted in Fig. 14 as a function of $E^{-1/2}$. This figure shows interesting features. We can first notice that the axial velocity damping times align on a single straight line for all Ekman numbers satisfying $E^{-1/2} < 45$. The damping time increases linearly with $E^{-1/2}$ as one could expect from classical Ekman layer scalings. The azimuthal velocity damping time has by contrast a totally different variation with respect to $E^{-1/2}$. For small $E^{-1/2}$ ($E^{-1/2} < 30$), that is close to threshold, it increases linearly with $E^{-1/2}$ as for the spin-up time, and is slightly larger than the axial velocity damping time. In this regime, we think that the spin-over mode is sufficiently small to have a negligible influence on the spin-down of the main rotation. Apparently, this is no longer the case for $E^{-1/2} > 30$: the main rotation becomes more rapidly damped than the spin-over mode. We believe that in this regime the damping of the main rotation is mostly due to an energy transfer towards the spin-over mode. The most amazing feature of these data is actually that in the regime $E^{-1/2} > 30$ the damping time of the main rotation decreases with respect to $E^{-1/2}$. This means that the faster is the rotation, the shorter its damping. We have no explanation for this unexpected result.

## 4. Conclusion

In this article, we have investigated the elliptical or tidal instability of a rotating fluid shell subjected to a stationary elliptic deformation. A model for the main flow has been first presented. Then its linear stability analysis has led to the determination of the growth rate of the spin-over mode of the rotating fluid shell as a function of the Ekman number. Visualization and anemometry measurements have illustrated the structure and the dynamics of this unstable mode. A good agreement between theory and experiments have been obtained, especially for the growth rate. The saturated nonlinear regime has also been analysed and appeared to be well-predicted by a simple nonlinear model for the spin-over mode amplitude. Finally, we have described the flow spin-down when the entrainment is abruptly stopped. Close to threshold, classical scalings in $E^{-1/2}$ for the damping times have been obtained. But strange and unexpected behavior have also been evidenced for smaller Ekman numbers.

The shell geometry is reminiscent of geophysical situations where the outer liquid cores of planets are elliptically deformed by tidal excitations. On Earth, it is believed (Kerswell, 1994) that the growth rate generated by the elliptic deformation is approximately balanced by the damping rate associated with Joule dissipation. There exist also other planets as Io where an elliptic instability could be possible and therefore play an important role in its core dynamics. Moreover, there probably exist exo-planets or other astrophysical binary systems in which such a dynamic might also be possible.